\documentclass[draft,tightenlines,nofootinbib,preprint,aps,eqsecnum,amsmath,amssymb]{revtex4}

\newcommand{\beq}{\begin{equation}}
\newcommand{\eeq}{\end{equation}}
\newcommand{\bea}{\begin{eqnarray}}
\newcommand{\eea}{\end{eqnarray}}

\newcommand{\sgn}{\epsilon}

\begin{document}

\title{The York Map and the Role of Non-Inertial Frames in the Geometrical View of the
Gravitational Field.}

\medskip

\author{Luca Lusanna}

\affiliation{ Sezione INFN di Firenze\\ Polo Scientifico\\ Via Sansone 1\\
50019 Sesto Fiorentino (FI), Italy\\  E-mail: lusanna@fi.infn.it}

\begin{abstract}

The role of non-inertial frames in a class of models of general
relativity is clarified by means of Dirac's theory of constraints.
The identification of a York canonical basis allows to give the
interpretation of the gauge variables as generalized inertial
effects and to identify the Dirac observables of the gravitational
field with generalized tidal effects. York time is the gauge
variable controlling the clock synchronization convention.
Differently from special relativity, the instantaneous 3-spaces are
dynamically determined.

\bigskip

Talk at the International School on {\it Astrophysical Relativity}
John Archibald Wheeler, Erice June 1-7, 2006.

\end{abstract}

\maketitle

\vfill\eject

Dirac's constraint theory \cite{1} is the natural language to
describe both gauge theories and gravitational physics. Even if the
Hamiltonian approach lies at the heart of the Faddeev-Popov measure
of the path integral and of the BRST method, both particle
physicists and general relativists tend to prefer the
configurational one due to its manifest covariance and to the
possibility of avoiding to face the problem of what is {\it time} in
a relativistic theory. However, this is an illusion, because the
problem cannot be eluded when we try to establish a well-posed
Cauchy problem for the field equations, in absence  of which we
cannot use the existence and uniqueness theorem for the solutions of
partial differential equations and speak of predictability. As a
consequence, the basic problem is how to separate the gauge degrees
of freedom from the gauge-invariant genuine dynamical variables (the
Dirac observables, DO). While the former are (modulo some
restriction) completely arbitrary, the latter have to satisfy
deterministic hyperbolic partial differential equations in every
completely fixed gauge. Now, only the Hamiltonian formalism has a
natural (even if still heuristic from the point of view of
mathematical rigor) tool to implement this separation: the
Shanmugadhasan canonical transformations \cite{2} adapted to the
first- and second-class constraints of the model and to the second
Noether theorem underlying their existence due to the local
invariances of the action functional. This type of transformations
were introduced by Dirac for the electromagnetic field in a seminal
paper \cite{3}, in which its DO were identified with the transverse
vector potential and electric field of the radiation gauge. This
work has been extended to Yang-Mills theory in Ref.\cite{4} and
applied in Ref. \cite{5} to the classical version of the standard
model of elementary particles. See Ref.\cite{6} for a review.

\medskip

However, these results {\it hold only in inertial frames in
Minkowski space-time} and are possible because gauge transformations
act on an {\it inner space}: the redundant gauge variables are
present only to implement the manifest covariance of the  model
under the action of the kinematical Poincare' transformations
connecting inertial frames and under local {\it inner} Lie groups.
As a consequence, in every formulation both at the classical and
quantum level the standard model of elementary particles and its
extensions are a chapter of the theory of representations of the
Poincare' group in inertial frames in Minkowski space-time (the
non-dynamical container of the fields).

\medskip

In special relativity the structure of the light-cones is an
absolute non-dynamical object \cite{7}: they are the  only
information (the conformal structure) given by the theory to an
(either inertial or accelerated) observer in each point of her/his
world-line. There is no notion of instantaneous 3-space, of spatial
distance and of one-way velocity of light between two observers
\footnote{By contrast in Newton physics there are distinct absolute
notions of {\it time} and {\it space}, so that we can speak of
absolute simultaneity and of instantaneous Euclidean 3-spaces with
the associated Euclidean spatial distance notion. This non-dynamical
chrono-geometrical structure is formalized in the so called Galilei
space-time. The Galilei relativity principle assumes the existence
of preferred inertial frames with inertial Cartesian coordinates
centered on inertial observers, connected by the kinematical group
of Galilei transformations. In Newton gravity the equivalence
principle states the equality of inertial and gravitational mass. In
non-inertial frames inertial (or fictitious) forces proportional to
the mass of the body appear in Newton's equations.}. The light
postulates say that the two-way (or round trip) velocity of light
$c$ (only one clock is needed in its definition) is constant and
isotropic. For an ideal inertial observer Einstein's convention for
the synchronization of distant clocks \footnote{The inertial
observer $A$ sends a ray of light at $x^o_i$ to a second accelerated
observer B, who reflects it towards A. The reflected ray is
reabsorbed by the inertial observer at $x^o_f$. The convention
states that the clock of B at the reflection point must be
synchronized with the clock of A when it signs ${1\over 2}\, (x^o_i
+ x^o_f)$.} selects the constant time hyper-planes of the inertial
frame having the observer as time axis as the instantaneous
Euclidean 3-spaces, with their Euclidean 3-geodesic spatial distance
and with the one-way and two-way velocities of light equal.
\medskip

But this convention does not work for realistic accelerated
observers, because coordinate singularities are produced in the
attempt (the 1+3 point of view) to build (either Fermi or rotating)
4-coordinates around the observer world-line. They must use the more
complex conventions arising from the introduction of an extra
structure: a global 3+1 splitting of Minkowski space-time (a choice
of {\it time}, the starting point of the Hamiltonian formalism).
Each space-like leaf of the associated foliation is {\it both a
Cauchy surface for the field equations and a convention} (different
from Einstein's one) {\it for clock synchronization}. If we
introduce {\it Lorentz-scalar observer-dependent radar
4-coordinates} $x^{\mu} \mapsto \sigma^A = (\tau ; \sigma^r)$, where
$x^{\mu}$ are Cartesian coordinates, $\tau$ is an arbitrary
monotonically increasing function of the proper time of the observer
and $\sigma^r$ curvilinear 3-coordinates having the observer
world-line as origin, this leads to the definition of a {\it
non-inertial frame} centered on the accelerated observer \cite{8}.
Every 3+1 splitting, satisfying certain M$\o$ller restrictions (to
avoid coordinate singularities) and with the leaves tending to
hyper-planes at spatial infinity (so that there are asymptotic
inertial observers to be identified with the fixed stars), gives a
conventional definition of instantaneous 3-space (in general a
Riemannian 3-manifold), of 3-geodesic spatial distance and of
one-way velocity of light (in general both point-dependent and
anisotropic). The inverse coordinate transformation $\sigma^A
\mapsto x^{\mu} = z^{\mu}(\tau ,\sigma^r)$ defines the {\it
embedding} of the simultaneity surfaces $\Sigma_{\tau}$ into
Minkowski space-time. The 3+1 splitting leads to the following
induced 4-metric (a functional of the embedding): ${}^4g_{AB}(\tau ,
\sigma^r) = {{\partial z^{\mu}(\sigma )}\over {\partial \sigma^A}}\,
{}^4\eta_{\mu\nu}\, {{\partial z^{\nu}(\sigma )}\over {\partial
\sigma^B}} = {}^4g_{AB}[z(\sigma )]$.

\medskip

Parametrized Minkowski theories \cite{9}, \cite{7} allow to give a
description of every isolated system (particles, strings, fields,
fluids), in which the transition from a 3+1 splitting to another one
(i.e. a change of clock synchronization convention) is a {\it gauge
transformation}. Given any isolated system admitting a Lagrangian
description, one makes the coupling of the system to an external
gravitational field and then replaces the 4-metric
${}^4g_{\mu\nu}(x)$ with the induced metric ${}^4g_{AB}[z(\tau
,\sigma^r)]$ associated to an arbitrary admissible 3+1 splitting.
The Lagrangian now depends not only on the matter configurational
variables but also on the embedding variables $z^{\mu}(\tau
,\sigma^r)$ (whose conjugate canonical momenta are denoted
$\rho_{\mu}(\tau ,\sigma^r)$). Since the action principle turns out
to be invariant under {\it frame-preserving diffeomorphisms}, at the
Hamiltonian level there are four first-class constraints ${\cal
H}_{\mu}(\tau ,\sigma^r) = \rho_{\mu}(\tau ,\sigma^r) - l_{\mu}(\tau
,\sigma^r)\, T^{\tau\tau}(\tau ,\sigma^r) - z^{\mu}_s(\tau
,\sigma^r)\, T^{\tau s}(\tau ,\sigma^r) \approx 0$ in strong
involution with respect to Poisson brackets, $\{ {\cal H}_{\mu}(\tau
,\sigma^r), {\cal H}_{\nu}(\tau ,\sigma_1^r)\} = 0$. Here
$l_{\mu}(\tau ,\sigma^r)$ are the covariant components of the unit
normal to $\Sigma_{\tau}$, while $z^{\mu}_s(\tau ,\sigma^r) =
{{\partial\, z^{\mu}(\tau ,\sigma^r)}\over {\partial\, \sigma^s}}$
are the components of three independent vectors tangent to
$\Sigma_{\tau}$. The quantities $T^{\tau\tau}$ and $T^{\tau s}$ are
the components of the energy-momentum tensor of the matter inside
$\Sigma_{\tau}$ describing its energy- and momentum- densities. As a
consequence, Dirac's theory of constraints implies that the
configuration variables $z^{\mu}(\tau ,\sigma^r)$ are arbitrary {\it
gauge variables}. Therefore, all the admissible 3+1 splittings,
namely all the admissible conventions for clock synchronization, and
all the admissible non-inertial frames centered on time-like
observers are {\it gauge equivalent}.\medskip

By adding four gauge-fixing constraints $\chi^{\mu}(\tau ,\sigma^r)
= z^{\mu}(\tau ,\sigma^r) - z^{\mu}_M(\tau ,\sigma^r) \approx 0$
($z^{\mu}_M(\tau ,\sigma^r)$ being an admissible embedding),
satisfying the orbit condition $det\, |\{\chi^{\mu}(\tau ,\sigma^r),
{\cal H}_{\nu}(\tau ,\sigma_1^r)| \not= 0$, we identify the
description of the system in the associated non-inertial frame
centered on a given time-like observer. The resulting effective
Hamiltonian for the $\tau$-evolution turns out to contain the
potentials of the {\it relativistic inertial forces} present in the
given non-inertial frame. As a consequence, the gauge variables
$z^{\mu}(\tau ,\sigma^r)$ describe the {\it spatio-temporal
appearances} of the phenomena in non-inertial frames, which, in
turn, are associated to {\it extended} physical laboratories using a
metrology for their measurements compatible with the notion of
simultaneity (distant clock synchronization convention) of the
non-inertial frame (think to the description of the Earth given by
GPS). Therefore, notwithstanding mathematics tends to use only
coordinate-independent notions, physical metrology forces us to
consider intrinsically coordinate-dependent quantities like the
non-inertial Hamiltonians. For instance, the motion of satellites
around the Earth is governed by a set of empirical coordinates
contained in the software of NASA computers: this is a {\it
metrological standard of space-time around the Earth}.
\medskip

Inertial frames centered on inertial observers are a special case of
gauge fixing in parametrized Minkowski theories. For each
configuration of an isolated system there is an special 3+1
splitting associated to it: the foliation with space-like
hyper-planes orthogonal to the conserved time-like 4-momentum of the
isolated system. This identifies an intrinsic inertial frame, the
{\it rest-frame}, centered on a suitable inertial observer (the
covariant non-canonical Fokker-Pryce center of inertia of the
isolated system) and allows to define the {\it Wigner-covariant
inertial rest-frame instant form of dynamics} for every isolated
system, which allows to give a new formulation of the relativistic
kinematics \cite{10} of N-body systems and continuous media
(relativistic centers of mass and canonical relative variables,
rotational kinematics and dynamical body frames, multipolar
expansions, M$\o$ller radius) and to find the theory of relativistic
orbits. Instead {\it non-inertial rest frames} are 3+1 splittings of
Minkowski space-time having the associated simultaneity 3-surfaces
tending to Wigner hyper-planes at spatial infinity.

\bigskip

Moreover it is now possible to define relativistic and
non-relativistic quantum mechanics of particles in non-inertial
frames \cite{11}, with a multi-temporal quantization scheme in which
the gauge variables $z^{\mu}(\tau ,\sigma^r)$ (the appearances) are
$c$-numbers (generalized times) and only the particle  degrees of
freedom are quantized. What is still lacking is the quantization of
a scalar field in non-inertial frames. Torre and Varadarajan
\cite{12} have shown that the traditional Tomonaga-Schwinger
approach does not lead in general to a unitary evolution. From the
3+1 point of view we have to restrict the 3+1 splittings to those
whose simultaneity leaves admit an instantaneous Fock space, in
which the Bogoliubov transformation between two such leaves of the
foliation is Hilbert-Schmidt (unitary evolution). Moreover all such
admissible 3+1 splittings must be unitarily gauge equivalent.
\bigskip

Things change dramatically when gravity is taken into account
\cite{7}. In general relativity there is no absolute notion: the
full chrono-geometrical structure of space-time is dynamical. The
relativistic description of gravity abandons the relativity
principle and replaces it with the equivalence principle. Special
relativity can be recovered only locally by a freely falling
observer in a neighborhood where tidal effects are negligible. As a
consequence, {\it global inertial frames do not exist}.
\medskip

The general covariance of Einstein's formulation of general
relativity leads to a type of gauge symmetry acting also on
space-time. The Hilbert action is invariant under coordinate
transformations ({\it passive} off-shell diffeomorphisms as local
Noether transformations), whereas the abstract differential
geometric formulation is invariant under {\it active}
diffeomorphisms of the space-time 4-manifold extended to tensors
(on-shell dynamical symmetries of Einstein's equations; it is
assumed that the space of solutions exists according to some notion
of integrability). Both in the off-shell and on-shell viewpoints the
gauge group is a group of diffeomorphisms acting also on space-time
(the same happens in every model with some type of reparametrization
invariance).

\medskip

Even if from a mathematical point of view the gauge variables are
still arbitrary degrees of freedom not determined by the field
equations, they are no more redundant variables of an inner space,
but are connected with the {\it appearances} of phenomena in the
various coordinate systems of Einstein's space-times.

\medskip

In Einstein's geometrical view of the gravitational field the basic
configuration variable is the metric tensor over space-time (10
fields), which, differently from every other field, has a double
role:

i) it is the mediator of the gravitational interaction, like every
other gauge field;

ii) it describes the dynamical chrono-geometrical structure of
space-time by means of the line element $ds^2 = {}^4g_{\mu\nu}(x)\,
dx^{\mu}\, dx^{\nu}$. As a consequence, it {\it teaches relativistic
causality} to the other fields: now the conformal structure (the
allowed  paths of light rays) is point-dependent.

\bigskip

In canonical ADM metric gravity \cite{13} (and in its tetrad gravity
extension \cite{14,15} needed for fermions  \footnote{This leads to
an interpretation of gravity based on a congruence of time-like
observers endowed with orthonormal tetrads: in each point of
space-time the time-like axis is the  unit 4-velocity of the
observer, while the spatial axes are a (gauge) convention for
observer's gyroscopes.}) we have again to start with the same
pattern of 3+1 splittings, to be able to define the Cauchy and
simultaneity surfaces for Einstein's equations. As a consequence,
and having in mind the inclusion of particle physics, we must select
a family of {\it non-compact} space-times $M^4$ with the following
properties:\hfill\break
 i) {\it globally hyperbolic} and {\it topologically trivial},
so that they can be foliated with space-like hyper-surfaces
$\Sigma_{\tau}$ diffeomorphic to $R^3$ (3+1 splitting of space-time
with $\tau$, the scalar parameter labeling the leaves, as a {\it
mathematical time});\hfill\break
 ii) {\it asymptotically flat at spatial infinity} and with
boundary conditions at spatial infinity independent from the
direction, so that the {\it spi group} of asymptotic symmetries is
reduced to the Poincare' group with the ADM Poincare' charges as
generators. In this way we can eliminate the {\it
super-translations}, namely the obstruction to define angular
momentum in general relativity, and we have the same type of
boundary conditions which are needed to get well defined non-Abelian
charges in Yang-Mills theory, opening the possibility of a unified
description of the four interactions with all the fields belonging
to same function space \cite{13}, \cite{6,7}. All these requirements
imply that the {\it admissible foliations} of space-time must have
the space-like hyper-surfaces tending in a direction-independent way
to Minkowski space-like hyper-planes at spatial infinity, which
moreover must be orthogonal there to the ADM 4-momentum. Therefore,
$M^4$ is {\it asymptotically Minkowskian} \cite{16} with the
asymptotic Minkowski metric playing the role of an {\it asymptotic
background}. Moreover the simultaneity 3-surfaces (the Riemannian
instantaneous 3-spaces) must admit an involution (Lichnerowicz
3-manifolds \cite{17}) allowing the definition of a generalized
Fourier transform with its associated concepts of positive and
negative energy, so to avoid the claimed impossibility to define
particles in curved space-times. \hfill\break
 iv) All the fields have to belong to suitable {\it weighted
Sobolev spaces} so that; i) the admissible space-like hyper-surfaces
are Riemannian 3-manifolds without asymptotically vanishing Killing
vectors \cite{16,18} (we furthermore assume the absence of any
Killing vector); ii) the inclusion of particle physics leads to a
formulation without Gribov ambiguity \cite{19},\cite{4}.
\medskip

In absence of matter the class of Christodoulou-Klainermann
space-times \cite{20}, admitting asymptotic ADM Poincare' charges
and an asymptotic flat metric meets these requirements.

\bigskip

This formulation, the {\it rest-frame instant form of metric and
tetrad gravity}, emphasizes the role of {\it non-inertial frames}
(the only ones existing in general relativity): each admissible 3+1
splitting identifies a global non-inertial frame centered on a
time-like  observer. In these space-times each simultaneity surface
is the rest frame of the 3-universe, there are asymptotic inertial
observers (the fixed stars) and the switching off of the Newton
constant in presence of matter leads to a deparametrization of these
models of general relativity to the non-inertial rest-frame instant
form of the same matter with the ADM Poincare' charges collapsing
into the usual kinematical Poincare' generators. This class of
space-times is suitable to describe the solar system (or the
galaxy), is compatible with particle physics and allows to avoid the
splitting of the metric into a background one plus a perturbation
\footnote{This splitting is the basic tool for the linearization of
Einstein's equations (see the theory of gravitational waves) and for
their replacement with a non-geometric spin-two theory over
Minkowski space-time, in which diffeomorphisms acting on space-time
are discontinuously replaced with gauge transformations acting on an
inner space. However, as shown by Deser \cite{21}, the
non-geometrical spin-two theory becomes inconsistent if we add the
energy-momentum tensor $T_{\mu\nu}$ of dynamical matter as a source:
the only way to recover consistency (at the price of loosing an
energy conservation law for the gravitational field) is to recover
Einstein's theory. Notwithstanding Deser's result, particle
physicists prefer to rely on Feynman's statement \cite{22} that {\it
the geometrical interpretation is not really necessary or essential
to physics}. The basic reason seems to be the absence of an energy
conservation law for the gravitational field, replaced by a
coordinate-dependent notion of energy density.}. With the addition
of suitable asymptotic terms it can probably be adapted to cosmology
\cite{23}.

\medskip

The first-class constraints of canonical gravity (8 in metric
gravity, 14 in tetrad  gravity \footnote{ Tetrad gravity has 10
primary first class constraints and 4 secondary first class ones.
Six of the primary constraints describe the extra freedom in the
choice of the tetrads. The other 4 primary (the vanishing of the
momenta of the lapse and shift functions) and 4 secondary (the
super-Hamiltonian and super-momentum constraints) constraints are
the same as in metric gravity. In Ref.\cite{14} 13 of the 14
constraints were solved: the super-Hamiltonian one can be solved
only after linearization \cite{15}.}) imply the existence of an
equal number of arbitrary gauge variables and of only 2+2 genuine
physical degrees of freedom of the gravitational field: $r_{\bar
a}(\tau ,\sigma^r)$, $\pi_{\bar a}(\tau ,\sigma^r)$. It can be shown
\cite{13,14,24} that the super-hamiltonian constraint generates
Hamiltonian gauge transformations implying the {\it gauge
equivalence} of clock synchronization conventions like it happens in
special relativity \footnote{The special relativistic constraints
${\cal H}_{\mu}(\tau ,\sigma^r) \approx 0$ are replaced by the
super-hamiltonian and super-momentum ones.} (no Wheeler-DeWitt
interpretation of it as a Hamiltonian). As shown in Refs.\cite{24}
\footnote{In these papers there is also a solution of Einstein's
Hole Argument: in this class of space-times it is possible to
identify the point-events of space-time by means of the four tidal
degrees of freedom of the gravitational field. In other words,
space-time and gravitational field are two faces of the same entity.
The previous identification is not valid in spatially compact
space-times without boundary, where the Dirac Hamiltonian weakly
vanishes and there is a frozen picture of dynamics.}, the gauge
variables describe {\it generalized inertial effects} (the
appearances), while the 2+2 gauge invariant DO describe {\it
generalized tidal effects}. In Refs.\cite{14,15} a Shanmugadhasan
canonical basis, adapted to 13 of the 14 tetrad gravity first class
constraints (not to the super-Hamiltonian one) was found. With its
help it can be shown \cite{24} that a completely fixed Hamiltonian
gauge is equivalent to the choice of a {\it non-inertial frame} with
its adapted radar coordinates centered on an accelerated observer
and its instantaneous 3-spaces (simultaneity surfaces): again this
corresponds to an extended physical laboratory  \footnote{Let us
remark that, if we look at Minkowski space-time as a special
solution of Einstein's equations with $r_{\bar a}(\tau ,\sigma^r) =
\pi_{\bar a}(\tau ,\sigma^r) = 0$ (zero Riemann tensor, no tidal
effects, only inertial effects), we find \cite{13} that the
dynamically admissible 3+1 splittings (non-inertial frames) must
have the simultaneity surfaces $\Sigma_{\tau}$ {\it 3-conformally
flat}, because the conditions $r_{\bar a}(\tau ,\sigma^r) =
\pi_{\bar a}(\tau ,\sigma^r) = 0$ imply the vanishing of the
Cotton-York tensor of $\Sigma_{\tau}$. Instead, in special
relativity, considered as an autonomous theory, all the non-inertial
frames compatible with the M$\o$ller conditions are admissible
\cite{8}, namely there is much more freedom in the conventions for
clock synchronization.}.

\medskip

In the rest-frame instant form of gravity \cite{13,14}, due to the
DeWitt surface term the effective Hamiltonian is not weakly zero
({\it no frozen picture} of dynamics), but is given by the weak ADM
energy $E_{ADM} = \int d^3\sigma\, {\cal E}_{ADM}(\tau ,\sigma^r)$
(it is the analogous of the definition of the electric charge as the
volume integral of the charge density in electromagnetism). The ADM
energy density depends on the gauge variables, namely it is a
coordinate-dependent quantity (the {\it problem of energy} in
general relativity). In a completely fixed gauge, in which the
inertial effects are given functions  of the DO, ${\cal
E}_{ADM}(\tau ,\sigma^r)$ becomes a well defined function only of
the DO's and there is a deterministic evolution of the DO's (the
tidal effects) given by the Hamilton  equations. A universe $M^4$ (a
4-geometry) is the equivalence class of all the completely fixed
gauges with gauge equivalent Cauchy data for the DO on the
associated Cauchy and simultaneity surfaces $\Sigma_{\tau}$. In each
completely fixed gauge (an off-shell non-inertial frame determined
by some set of gauge-fixing constraints determining the gauge
variables in terms of the tidal ones) we find the solution for the
DO in that gauge (the tidal effects) and then the explicit form of
the gauge variables (the inertial effects). As a consequence, the
final admissible (on-shell gauge equivalent) non-inertial frames
associated to a 4-geometry (and their instantaneous 3-spaces, i.e.
their clock synchronization conventions) are {\it dynamically
determined} \cite{24}.

\bigskip

A first application of this formalism  has been the determination
\cite{15} of {\it post-Minkowskian background-independent
gravitational waves} in a completely fixed non-harmonic 3-orthogonal
gauge with diagonal 3-metric. It can be shown that the requirements
$r_{\bar a}(\tau ,\sigma^r) << 1$, $\pi_{\bar a}(\tau ,\sigma^r) <<
1$ lead to a weak field approximation based on a Hamiltonian
linearization scheme: i) linearize the Lichnerowicz equation (i.e.
the super-Hamiltonian constraint), determine the conformal factor of
the 3-metric and then the lapse and shift functions; ii) find
$E_{ADM}$ in this gauge and disregard all the terms more than
quadratic in the DO; iii) solve the Hamilton equations for the DO.
In this way we get a solution of linearized Einstein's equations, in
which the configurational DO $r_{\bar a}(\tau ,\sigma^r)$ play the
role of the {\it two polarizations} of the gravitational wave.

\bigskip

In Refs.\cite{25,26} there is the description of relativistic fluids
and of the Klein-Gordon field in the framework of parametrized
Minkowski theories. This formalism allows to get the Lagrangian of
these matter systems in the  formulation of tetrad gravity of
Refs.\cite{14,15}. The resulting first-class constraints depend only
on the mass density ${\cal M}(\tau ,\sigma^r)$ (which is
metric-dependent) and the mass-current density ${\cal M}_r(\tau
,\sigma^r)$ (which is metric-independent) of the matter. For Dirac
fields the situation is more complicated due to the presence of
second class constraints (see Ref.\cite{27} for the case of
parametrized Minkowski theories with fermions). It turns out that
the point Shanmugadhasan  canonical transformation of Ref.\cite{15},
adapted to 13 of the 14 first class constraints is not suited for
the inclusion of matter due to its {\it non-locality}. Therefore the
search started  for a local point Shanmugadhasan transformation
adapted only to 10 of the 14 constraints, i.e. not adapted to the
super-Hamiltonian and super-momentum constraints.

\medskip

The new insight came from the so-called York - Lichnerowicz
conformal approach \cite{28,29,30,31}  to metric gravity in globally
hyperbolic ({\it but spatially compact}) space-times. The starting
point is the decomposition ${}^3g_{ij} = \phi^4\, {}^3{\hat g}_{ij}$
of the 3-metric on an instantaneous 3-space $\Sigma_o$ of a 3+1
splitting of space-time in the product of a {\it conformal factor}
$\phi = (det\, {}^3g)^{1/12}$ and a {\it conformal 3-metric}
${}^3{\hat g}_{ij}$ with $det\, {}^3{\hat g}_{ij} = 1$ (${}^3{\hat
g}_{ij}$ contains 5 of the 6 degrees of freedom of ${}^3g_{ij}$).
The extrinsic curvature 3-tensor ${}^3K_{ij}$ of $\Sigma_o$  is
decomposed in its trace ${}^3K$ (the {\it York time}) plus the {\it
distorsion tensor}, which is the sum of a TT \footnote{Traceless and
transverse with respect to the conformal 3-metric.} symmetric
2-tensor ${}^3A_{ij}$ (2 degrees of freedom) plus the 3-tensor
${}^3W_{i;j} + {}^3W_{j;i} - {2\over 3}\, {}^3g_{ij}\,
{}^3W^k{}_{;k}$ depending on a covariant 3-vector ${}^3W_i$ ({\it
York gravitomagnetic vector potential}; 3 degrees of freedom).
Having fixed the lapse and shift functions of the 3+1 splitting and
having put ${}^3K = const.$, one assigns ${}^3{\hat g}_{ij}$ and
${}^3A_{ij}$ on the Cauchy surface $\Sigma_o$. Then, ${}^3W_i$ is
determined by the super-momentum constraints on $\Sigma_o$ and
$\phi$ is determined by the super-Hamiltonian constraint on
$\Sigma_o$. Then, the remaining Einstein's equations (see
Refs.\cite{18,28} for the existence and unicity of solutions)
determine the time derivatives of ${}^3g_{ij}$ and of ${}^3K_{ij}$,
allowing to find the time development from the initial data on
$\Sigma_o$. However, a canonical basis adapted the the previous
splittings was never found. The only result is contained in
Ref.\cite{32}, where it was shown that, having fixed ${}^3K$, the
transition from the non-canonical variables ${}^3{\hat g}_{ij}$,
${}^3A_{ij}$, ${}^3W_i$ to the space of the gravitational initial
data satisfying the constraints is a canonical transformation, named
{\it York map}.\bigskip

In Ref.\cite{33}  a new parametrization of the original 3-metric
${}^3g_{ij}$ was proposed, which allows to find local point
Shanmugadhasan canonical transformation, adapted to 10 of the 14
constraints of tetrad gravity, implementing a York map. The 3-metric
${}^3g_{rs}$ may be diagonalized with an {\it orthogonal} matrix
$V(\theta^r)$, $V^{-1} = V^T$, $det\, V = 1$, depending on 3 Euler
angles $\theta^r$. The gauge Euler angles $\theta^r$ give a
description of the 3-coordinate systems on $\Sigma_{\tau}$ from a
local point of view, because they give the orientation of the
tangents to the 3 coordinate lines through each point (their
conjugate momenta $\pi_i^{(\theta )}$ are determined by the
super-momentum constraints and replace the York gravitomagnetic
potential ${}^3W_i$), $\phi$ is the conformal factor of the
3-metric, i.e. the unknown in the super-hamiltonian constraint
\footnote{The only role of the conformal decomposition ${}^3g_{ij} =
\phi^4\, {}^3{\hat g}_{ij}$ is to identify the conformal factor
$\phi$ as the natural unknown in the super-Hamiltonian constraint,
which becomes the {\it Lichnerowicz equation}. See Ref.\cite{13} for
a different justification of this result based on constraint theory
and the two notions of strong and weak ADM energy.} (its conjugate
momentum is the gauge variable describing the form of the
simultaneity surfaces $\Sigma_{\tau}$), while the two independent
eigenvalues of the conformal 3-metric ${}^3{\hat g}_{rs}$ (with
determinant equal to 1) describe the genuine {\it tidal} effects
$R_{\bar a}$,  $\bar a = 1,2$, of general relativity (the non-linear
"graviton", with conjugate momenta $\Pi_{\bar a}$). In the York
canonical basis \cite{33} the gauge variable, which describes the
freedom in the choice of the clock synchronization convention, i.e.
in the definition of the instantaneous 3-spaces $\Sigma_{\tau}$, is
the trace ${}^3K(\tau ,\sigma^r)$ of the extrinsic curvature of
$\Sigma_{\tau}$. It is both the York time and the momentum conjugate
to the conformal factor.

\medskip

The tidal effects $R_{\bar a}(\tau ,\sigma^r)$, $\Pi_{\bar a}(\tau
,\sigma^r)$, are DO {\it only} with respect to the gauge
transformations generated by 10 of the 14 first class constraints.
Let us remark that, if we fix completely the gauge and we go to
Dirac brackets, then the only surviving dynamical variables $R_{\bar
a}$ and $\Pi_{\bar a}$ become two pairs of {\it non canonical} DO
for that gauge: the two pairs of canonical DO have to be found as a
Darboux basis of the copy of the reduced phase space identified by
the gauge and they will be (in general non-local) functionals of the
$R_{\bar a}$, $\Pi_{\bar a}$ variables. This shows the importance of
canonical bases like the York one: the tidal effects are described
by {\it local} functions of the 3-metric and its conjugate
momenta.\medskip

Since the conformal factor $\phi$  and and the momenta
$\pi_i^{(\theta )}$ conjugate to the Euler angles $\theta^r$  are
determined by the super-Hamiltonian and super-momentum constraints,
the {\it arbitrary gauge variables} of the York canonical basis are
$\alpha_{(a)}$, $\varphi_{(a)}$, $\theta^i$, $\pi_{\tilde \phi}$,
$n$ and ${\bar n}_{(a)}$. As shown in Refs.\cite{24,33}, they
describe the following generalized {\it inertial effects}:

a) the angles $\alpha_{(a)}(\tau ,\sigma^r)$ and the boost
parameters $\varphi_{(a)}(\tau ,\sigma^r)$  describe the
arbitrariness in the choice of a tetrad to be associated to a
time-like observer, whose world-line goes through the point $(\tau
,\sigma^r)$. They fix {\it the unit 4-velocity of the observer and
the conventions for the gyroscopes and their transport along the
world-line of the observer}.

b) the angles $\theta^i(\tau ,\sigma^r)$ (depending only on the
3-metric) describe the arbitrariness in the choice of the
3-coordinates on the simultaneity surfaces $\Sigma_{\tau}$ of the
chosen non-inertial frame  centered on an arbitrary time-like
observer. Their choice induces a pattern of {\it relativistic
standard inertial forces} (centrifugal, Coriolis,...), whose
potentials are contained in the weak ADM energy $E_{ADM}$. These
inertial effects are the relativistic counterpart of the
non-relativistic ones (they are present also in the non-inertial
frames of Minkowski space-time).

c) the {\it shift} functions ${\bar n}_{(a)}(\tau ,\sigma^r )$,
appearing in the Dirac Hamiltonian, describe which points on
different simultaneity surfaces have the same numerical value of the
3-coordinates. They are the inertial potentials describing the
effects of the non-vanishing off-diagonal components ${}^4g_{\tau
r}(\tau ,\sigma^r)$ of the 4-metric, namely they are the {\it
gravito-magnetic potentials} \footnote{In the Post-Newtonian
approximation in {\it harmonic gauges} they are the counterpart of
the electro-magnetic vector potentials describing magnetic fields
\cite{31}, \cite{15}: A) $N = 1 + n$, $n\, {\buildrel {def}\over
=}\, - {{4\, \sgn}\over {c^2}}\, \Phi_G$ with $\Phi_G$ the {\it
gravito-electric potential}; B) $n_r\, {\buildrel {def}\over =}\,
{{2\, \sgn}\over {c^2}}\, A_{G\, r}$ with $A_{G\, r}$ the {\it
gravito-magnetic} potential; C) $E_{G\, r} =
\partial_r\, \Phi_G - \partial_{\tau}\, ({1\over 2}\, A_{G\, r})$ (the {\it
gravito-electric field}) and $B_{G\, r} = \epsilon_{ruv}\,
\partial_u\, A_{G\, v} = c\, \Omega_{G\, r}$ (the {\it
gravito-magnetic field}). Let us remark that in arbitrary gauges the
analogy with electro-magnetism  breaks down.} responsible of effects
like the dragging of inertial frames (Lens-Thirring effect)
\cite{31} in the post-Newtonian approximation.

d) $\pi_{\phi}(\tau ,\sigma^r )$, i.e. the York time ${}^3K(\tau
,\sigma^r)$, describes the arbitrariness in the shape of the
simultaneity surfaces $\Sigma_{\tau}$ of the non-inertial frame,
namely the arbitrariness in the choice of the convention for the
synchronization of distant clocks. Since this variable is present in
the Dirac Hamiltonian, it is a {\it new inertial potential}
connected to the problem of the relativistic freedom in the choice
of the {\it instantaneous 3-space}, which has no non-relativistic
analogue (in Galilei space-time time is absolute and there is an
absolute notion of Euclidean 3-space). Its effects are completely
unexplored. For instance, since the sign of the trace of the
extrinsic curvature may change from a region to another one on the
simultaneity surface $\Sigma_{\tau}$, {\it the associated inertial
force in the Hamilton equations may change from attractive to
repulsive in different regions}.

e) the {\it lapse} function $N(\tau ,\sigma^r) = 1 + n(\tau ,
\sigma^r )$, the lapse function appearing in the Dirac Hamiltonian,
describes the arbitrariness in the choice of the unit of proper time
in each point of the simultaneity surfaces $\Sigma_{\tau}$, namely
how these surfaces are packed in the 3+1 splitting.
\medskip

As a consequence, differently from special relativity, the
conventions for clock synchronization and the whole
chrono-geometrical structure of $M^4$ (gravito-magnetism, 3-geodesic
spatial distance on $\Sigma_{\tau}$, trajectories of light rays in
each point of $M^4$, one-way velocity of light) are {\it dynamically
determined } \cite{24}.

\medskip

The use of Dirac theory of constraints introduces a different point
of view on the gauge-fixing and the Cauchy problem. While the gauge
fixing to the extra 6 primary constraints fixes the tetrads (i.e.
the spatial gyroscopes and their transport law), the gauge fixing to
the 4 primary plus 4 secondary constraints follows a different
scheme from the one used in the York-Lichnerowicz approach, which
influenced contemporary numerical gravity. Firstly one adds the 4
gauge fixings to the secondary constraints (the super-Hamiltonian
and super-momentum ones), i.e. one fixes ${}^3K$, i.e.the
simultaneity 3-surface, and the 3-coordinates on it (namely 3 of the
5 degrees of freedom of the conformal 3-metric ${}^3{\hat g}_{ij}$).
The preservation in time of these 4 gauge fixings generates other 4
gauge fixing constraints determining the lapse and shift functions
consistently with the shape of the simultaneity 3-surface and with
the choice of 3-coordinates on it (here is the main difference with
the conformal approach and most of the approaches to numerical
gravity).

\bigskip

The clarification of the interpretational issues allowed by the York
canonical basis will allow to face many problems.\medskip

1) To understand better the Hamiltonian distinction between inertial
and tidal effects, a detailed study of the Post-Newtonian solutions
of Einstein's equations adopted by the IAU conventions \cite{34} for
the barycentric and geocentric celestial reference frames has begun.
In particular it will clarify the mixing of the general relativistic
effects like the gravitational redshift with the special
relativistic ones like the Doppler effect and the Coriolis forces
near the rotating Earth. This is no more an academic research,
because in a few years the European Space Agency (ESA) will start
the mission ACES \cite{35} about the synchronization of a
high-precision laser-cooled atomic clock on the space station with
similar clocks on the Earth surface by means of microwave signals.
If the accuracy of 5 picosec. will be achieved, it will be possible
to make a coordinate-dependent test of effects at the order $1/c^3$,
like the second order Sagnac effect (sensible to Earth rotational
acceleration) and the general relativistic Shapiro time-delay
created by the geoid \cite{36}. It will be important to find the
Post-Newtonian deviation from Einstein's convention to be able to
synchronize two such clocks and to understand which metrological
protocols have to be used for time dissemination at this level of
accuracy. Therefore, the problem of clock synchronization is
becoming every day more important due to GPS, to the ACES mission of
ESA, to the Bepi-Colombo mission to Mercury and to the future space
navigation inside the solar system.

\medskip

2) The geometric vision of space-time will soon be enriched with the
Hamiltonian reformulation of the Newman-Penrose formalism, in
particular of the 10 Weyl scalars. This will allow a) to search the
Bergmann observables \cite{24} (special DO describing  {\it scalar
tidal effects}) and try to understand which inertial effects may
have a coordinate-independent form (like gravito-magnetism) and
which are intrinsically coordinate-dependent like the ADM energy
density; b) to look for the existence of a closed Poisson algebra of
scalars and for Shanmugadhasan canonical bases incorporating the
Bergmann observables, to be used to find new expressions for the
super-hamiltonian and super-momentum constraints, hopefully easier
to be solved. If the Torre-Varadarajan no-go theorem \cite{12} can
be avoided and the scalar field can be quantized in an admissible
set of non-inertial frames in Minkowski space-time, it will be
possible to arrive at a multi-temporal background- and coordinate-
independent multi-temporal quantization (see Ref.\cite{11}) of the
gravitational field, in which only the Bergmann observables (the
scalar tidal effects) are quantized.\medskip

\bigskip

3) The study the 2-body problem in general relativity in various
coordinate systems at least in the weak field approximation, with a
Grassmann regularization of the self-energies, following the track
of Refs.\cite{37} is now possible. In these papers the use of
Grassmann-valued electric charges to regularize the Coulomb
self-energies allowed to arrive to the Darwin and Salpeter
potentials starting from classical electrodynamics of scalar and
spinning particles, instead of deriving them from quantum field
theory. The solution of the Lichnerowicz equation would allow to
find the expression of the relativistic Newton and gravito-magnetic
action-at-a-distance potentials between the two bodies (sources,
among other effects, of the Newtonian tidal effects) and the
coupling of the particles to the DO of the gravitational field (the
genuine tidal effects) in various radar coordinate systems: it would
amount to a re-summation of the $1/c$ expansions of the
Post-Newtonian approximation. Also the relativistic version of the
quadrupole formula for the emission of gravitational waves from the
binary system could be obtained and some  understanding of how is
distributed the gravitational energy in different coordinate systems
could be  obtained \footnote{If the ADM energy-momentum pseudo-
tensor will be identified, its reformulation as the energy-momentum
tensor of a viscous pseudo-fluid will allow to check whether the
pressure field is positive-definite or not, namely whether the
gravitational energy contributes to dark energy.}. It would also be
possible to study the deviations induced by Einstein's theory from
the Keplerian standards for problems like the radiation curves of
galaxies, whose Keplerian interpretation implies the existence of
dark matter \footnote{Since the relativistic inertial forces are
present in the Hamilton equations for the gravitational DO and for
the matter in the gauge-dependent instantaneous 3-space, they may be
a relativistic alternative to the MOND model \cite{38} (modification
of the non-relativistic Newton equations on the acceleration side
for slow accelerations). At least part of the dark mattr could be
explained by relativistic inertial effects. See Ref.\cite{39} for a
possible gravito-magnetic origin of dark matter}.\medskip

4) With more general types of matter (fluids, electro-magnetic
field) we could define Hamiltonian numerical gravity (for instance
with a post-Minkowskian development in powers of the Newton
constant) and try to find strong-field approximations to be used in
the gravitational collapse and to find the strong-field deviations
from the Newton potential. This last problem is completely open in
every approach.

\end{document}